\documentstyle{lamuphys}
\def\be{\begin{equation}}
\def\ee{\end{equation}}
\def\benn{$$}
\def\eenn{$$}
\def\bea{\begin{eqnarray}}
\def\eea{\end{eqnarray}}
\def\beann{\begin{eqnarray*}}
\def\eeann{\end{eqnarray*}}

\newcommand{\ket}[1]{|\kern.3ex#1\kern.3ex\rangle}
\newcommand{\bra}[1]{\langle\kern.3ex #1 \kern.3ex|}
\newcommand{\APPROX}[1]{                
   {{\raisebox{-.3cm}{$\textstyle\simeq$}} \atop {\scriptstyle{#1}}}}
\newcommand{\APPROXs}[1]{               
   {{\raisebox{-.3cm}{$\textstyle\sim$}} \atop {\scriptstyle{#1}}}}
\newcommand{\EXP}[1]{{\mbox{\large e}}^{#1}}
\newcommand{\mean}[1]{\left\langle #1 \right\rangle} 

\newcommand\Dslash{D\hspace{-0.25cm}/}               
\renewcommand{\log}{\mathop{\mathrm{ln}}\nolimits}

\def\D{{\rm d}}
\def\I{{\rm i}}

\begin{document}

\setcounter{page}{313}


\title{One-Dimensional Disordered Supersymmetric 
      Quantum Mechanics: A Brief Survey}

\author{Alain Comtet\inst{1,2}
    and Christophe Texier\inst{1}}

\institute{Division de Physique Th\'eorique\ddag, IPN B\^at. 100, 91406 Orsay 
C\'edex, France.
  \and     L.P.T.P.E, Universit\'e Paris 6, 4 place Jussieu, 75252 Paris
C\'edex 05, France.}

\maketitle

\noindent
E-mail: comtet@ipno.in2p3.fr\\
E-mail: texier@ipno.in2p3.fr\\
\ddag\hspace{0cm} Unit\'e de recherche des Universit\'es Paris 11 et Paris 6 
associ\'ee au CNRS.

\begin{abstract}
We consider a one-dimensional model of localization based on the Witten
Hamiltonian of supersymmetric quantum mechanics. The low energy spectral
properties are reviewed and compared with those of other models with 
off-diagonal disorder. Using recent results on exponential functionals 
of a Brownian motion we discuss the statistical properties of the ground 
state wave function and their multifractal behaviour.
\end{abstract}

\bigskip

\noindent IPNO/TH 97-21

\bigskip

\markboth{Alain Comtet and Christophe Texier}
         {One-Dimensional Disordered Supersymmetric Quantum Mechanics}
\pagestyle{myheadings}


\section{Introduction}

Many physical systems can be described using the concept of a random
Hamiltonian. Such a formulation is useful when the Hamiltonian depends on a
set of quenched variables. In most cases it is convenient to describe them
by a set of random variables distributed according to some probability law.
Consider, for instance, the quantum dynamics of a particle interacting with
randomly distributed scatterers. If the potential is a sum of two-body
potentials the Hamiltonian may be taken in the form
\benn
H=\frac{\vec p^2}{2m} + \sum_{k=1}^N V(\vec r-\vec r_k)
.\eenn
Here the quenched variables are the positions of the impurities $\vec r_k$
and the number $N$ of scatterers. They can, for instance, be modelled by
independently random variables distributed according to a Poisson distribution.

In this context one of the most elementary quantities of physical interest is
the average density of states $\rho(E)$. If the potential is
repulsive and short range then $\rho(E)$ vanishes exponentially
at the bottom of the spectrum
\benn
\rho(E)\APPROXs{E\to 0}\EXP{-C\,E^{-d/2}}
\eenn
where $d$ is the dimension of the space.
This non-analytic behaviour was first discussed by Lifshits
(\cite{lifshits}) and then studied by a number of authors (see for example 
\cite{luttinger}). The
physical mechanism which leads to this behaviour is the occurence of large 
regions of space that are free from impurities and where the particle can move
freely. Although these are exponentially rare events, they nevertheless
contribute in the thermodynamic limit. This singular behaviour may also be
derived using instanton techniques (\cite{neuberger}).

However there do exist some systems for which the density of states has 
a very different behaviour:\\
\begin{enumerate}

\item The vibrations of a chain consisting of harmonic strings and random
masses gives a spectral density with an accumulation of states at low
frequency. This model, first introduced by Dyson (\cite{dyson}), is in fact
equivalent to the one-dimensional Anderson model with off-diagonal disorder.

\item In particle physics, the investigation of the random Dirac operator
has stimulated interesting conjectures related to chiral symmetry breaking
(\cite{stern}).
A particular model in dimension $2+1$ is to take as a Hamiltonian the square
of the Euclidean Dirac operator coupled to a random magnetic field
$H=-\Dslash^2=-(\partial_\mu +\I A_\mu)^2+\frac{1}{2}\sigma_{\mu\nu}F^{\mu\nu}$.
In this case the low energy density of states must fullfil the inequality
(\cite{casherneu}) $\rho(E)>\rho_o(E)$, where $\rho_o(E)$ is the 
free density of states. 
This means that there is an accumulation of low energy states
which contributes to the chiral condensate.

\item The one-dimensional Schr\"odinger Hamiltonian
$H=-\frac{\D^2}{\D x^2} + \phi^2(x) + \sigma_3\phi'(x)$
which was first introduced by Witten (\cite{witten}) as a toy model of 
supersymmetric quantum mechanics (for a review see for example \cite{cooper}) 
provides a localization model with very
unusual low energy spectral properties. In certain cases it gives rise to an
accumulation of levels at low energy. The density of states displays either
a power law behaviour $\rho(E)\APPROXs{E\to 0}E^{\mu-1}$ or a
logarithmic singularity
$\rho(E)\APPROXs{E\to 0}\frac{1}{E|\log E|^3}$
of the same form as in the Dyson model (\cite{dyson}).

\end{enumerate}

A common feature of these three models is the fact that the zero energy wave
function is exactly known for any realization of the disorder. Using this
property we have presented, for the supersymmetric model (\cite{locprop}),
a physical
picture that accounts for the different behaviours of the density of states
at the bottom of the spectrum. We believe that this model is probably
generic, by which we mean representative of a whole class of systems in which
the disorder is encoded in the ground state. Since this 
model is easier to handle, because one can use a wealth of techniques
specific to one-dimensional systems, we will concentrate on this case. The
recent literature shows a revival of interest for these problems, mainly in
the context of condensed matter physics. We will briefly comment on this
work and also draw attention to earlier work which is scattered in the
literature and has so far remained unnoticed.

In part \ref{dos} we will review the basic mechanism which leads to these
singularities and underline the differences with the usual Lifshits
singularities. We will also mention some recent applications of the
supersymmetric model to quantum spin chains. Applications to classical
diffusion in a random medium (\cite{BCGD}, \cite{classdiff}, \cite{oshanin2}) 
will not be discussed here.

In part \ref{correlation}, following (\cite{broderix}) and (\cite{tsvelik}) we
compute the correlation function of the zero energy states.

In part \ref{moments} we discuss the fluctuation properties of the ground
state wave function using two different approaches. When $\phi(x)$ is
white noise the wave function is an exponential functional of the Brownian
motion. Such functionals were studied extensively both in the mathematical
(\cite{yor3}) and physical literature (\cite{fluxdist}, \cite{expfunc}). 
We will use our previous work to compute their statistical properties.


\section{Spectral properties}\label{dos}

The one-dimensional Schr\"odinger Hamiltonians
\be\label{susyham}
H_{\pm}=-\frac{\D^2}{\D x^2} +\phi^2(x) \pm\phi'(x)
\ee
may be rewritten in the factorized forms $H_{+}=Q^\dagger Q$ and
$H_{-}=Q Q^\dagger$, where $Q\equiv-\frac{\D}{\D x}+\phi(x)$. This implies that
$H_+$ and $H_-$ have the same spectrum for $E>0$. When $\phi(x)$ is random,
they are characterized  by the same  localization length and density of states.
These quantities have been computed exactly in two cases for which we now
recall the main results.

\subsection{White noise potential}

\benn
\left\{\begin{array}{l}
  \mean{\phi(x)}=F_0\\
  \mean{\phi(x)\phi(x')}-\mean{\phi(x)}^2=\sigma\,\delta(x-x')
\end{array}\right.
\eenn
The integrated density of states $N(E)$ and the localization length
$\lambda(E)$ are respectively
\bea\label{ovch}
&&N(E)=\frac{2\sigma}{\pi^2}\frac{1}{J^2_\mu(z)+N^2_\mu(z)}\\
&&\lambda^{-1}(E)=
-\frac{\sigma z}{2}\frac{\D}{\D z}\log{\left(J^2_\mu(z)+N^2_\mu(z)\right)}
\nonumber
\eea
where $z\equiv\frac{\sqrt E}{\sigma}$ and $\mu\equiv\frac{F_0}{\sigma}$.
$J_\mu(z)$ and $N_\mu(z)$ are Bessel functions.

Equation (\ref{ovch}) was first obtained by Ovchinnikov and Erikmann 
(\cite{ovchinni}) and then rediscovered independently in (\cite{BCGD2}).
These results can be derived either by the node counting method 
(\cite{luck}) or by the
replica trick (\cite{BCGD}). By this latter approach one can also compute 
the Green's function at non-coinciding points.

The low energy behaviour of the density of states and the localization length
are given for $\mu=0$ by
\beann
&& N(E)       \APPROXs{E\to 0}\frac{1}{\log^2 E}\\
&& \lambda(E) \APPROXs{E\to 0}-\log E
\eeann
and by
\beann
&& N(E)       \APPROXs{E\to 0} E^\mu \\
&& \lambda(E) \APPROXs{E\to 0} E^{\mu-1}
\eeann
when $\mu\neq0$.

\subsection{Random telegraph process}

The function $\phi(x)$ is described by an ensemble of rectangular barriers 
(\cite{pastur})  with
alternating heights $\phi_0$ and $\phi_1$ of random length $l$ distributed
according to an exponential law $p_{0,1}(l)=n_{0,1}\EXP{-n_{0,1}l}$.

In the case $\phi_0=-\phi_1$ this model yields the same low energy behaviour
as above (\cite{locprop}). 
The parameter $\mu$ is now given by $\mu=\frac{n_1-n_0}{2\phi_0}$. The
main interest of this model is to provide a physical picture of the low
energy behaviour.

For $\mu\neq 0$ the potential $V(x)=\phi^2+\phi'$ is constant everywhere
except at the positions where $\phi(x)$ has a discontinuity. One thus
obtains a sequence of $\delta$ functions with alternating signs. The
attractive $\delta$ potentials can support bound states which would have
exactly
zero energy if one would ignore the couplings to the other peaks. By taking
carefully into account these couplings, one can recover the low energy
power law behaviour (\cite{locprop}). The physical picture that emerges from 
this analysis is that the 
{\em low energy states are localized at the positions of the
impurities}. Therefore this is just the opposite mechanism from that in the
Lifshits case. It would be interesting to generalize this approach to higher
dimensions where similar behaviour can also occur. 

For $\mu=0$, since the positive and negative $\delta$ functions play a
symmetric role this argument doesn't apply anymore. A study of the low
energy states on a finite interval shows that the existence of quasi
zero-modes can account for the logarithmic behaviour of the density of states.

Imposing Dirichlet boundary conditions on a finite interval $[-R,R]$ one 
finds that the ground state energy  is (\cite{locprop}, \cite{gsenergy})
\benn
E_0(R,\{\phi\})\simeq
\frac{1}{\int_{-R}^R \D x'\,\psi_0^2(x')} \left(
  \frac{1}{\int_{-R}^0\frac{\D x}{\psi_0^2(x)}}
+ \frac{1}{\int_0^R \frac{\D x}{\psi_0^2(x)}}
\right)
.\eenn
This result is obtained by approximating the true ground state wave function
near the boundaries by a suitable linear combination of the two linearly
independent solutions of $H_+\psi=0$:
\be
\psi_0(x)=\EXP{\int^x \D x'\,\phi(x')}
\ee
and
\be
\psi_1(x)=\psi_0(x)\int^x\frac{\D x'}{\psi_0^2(x')}
.\ee

If $\phi(x)$ is a white noise or a random telegraph process with Poissonian
lengths, the typical behaviour of $\psi_0(x)$, given by the central limit
theorem, is $\psi_0(x)\sim\EXP{\pm\sqrt{x}}$.
Replacing $\psi_0(x)$ by its typical behaviour one finds that the energy
$E_0$ is exponentially small in the length of the system
$E_0\sim\EXP{-\sqrt{R}}$.
Therefore a quasi zero mode of energy $E$ has a typical spatial extension
$2R$ such that $R\sim\log^2 E$.
Coming back to the whole line one finds that the number of such states per
unit length is
\benn
N(E)=\frac{1}{2R}\sim\frac{1}{\log^2 E}
.\eenn
Obviously this argument may be generalized to any one-dimensional disordered
system for which the zero energy wave functions can be expressed in terms of
the potential. This is in particular the case of the Anderson model with
off-diagonal disorder. The discrete Schr\"odinger equation may be written in
the form
\be\label{recu}
\beta_{n+1}\varphi_{n+1}+\beta_n\varphi_{n-1}=E\varphi_n
\ee
where $\beta_n$ are random variables. The model of Dyson of an harmonic
chain with random masses belongs to this class. For any configuration of the
disorder one can write down two independent zero energy solutions. One of
them is obtained by solving the recurrence relation (\ref{recu}) with $E=0$.
A zero energy state satisfying the boundary conditions $\varphi_0=1$ and 
$\varphi_1=0$ is
\benn
\varphi_{2n}=\prod_{k=1}^n\frac{\beta_{2k-1}}{\beta_{2k}}
.\eenn
If the $\beta_n$ are independent identically distributed random variables,
the typical behaviour of $\varphi_n$, given by the central limit theorem, is
again of the form $|\varphi_n|\sim\EXP{\pm\sqrt{n}}$. One will therefore get
the same low energy behaviour as before. For earlier references see
(\cite{theodorou}, \cite{markos}, \cite{bovier}); another derivation of the 
logarithmic singularity is given in (\cite{eggarter}).

\subsection{Remarks}

\begin{enumerate}
\item All these arguments are based on typical realizations of  disorder.
There exist however certain quantities whose behaviour cannot be obtained by
this type of reasoning. This is in particular the case of the average ground
state energy. Its dependence on the size of the sample has been obtained in
(\cite{gsenergy}). It is given by the stretched-exponential function
$\mean{E_0}=\exp{(-R^\xi)}$ where the exponent $\xi$ depends only on the
nature of the correlations in the potential (the Gaussian white noise
corresponds to $\xi=1/3$). In (\cite{gsenergy}) it is shown that this
behaviour is indeed supported by atypical realizations of the random
potential. 

\item The existence of a singular behaviour in the density of states
implies, by the Thouless formula, that there will be a corresponding
singularity in the localization length $\lambda(E)$. This quantity indeed 
diverges as $\log E$ which reflects the appearance of a critical state at 
$E=0$. It was recently pointed out (\cite{steiner}, \cite{balents}) 
that there exists another length scale in the system - the correlation length
which controls the decay of the two-point Green's function. It behaves like 
$\log^2E$ and thus
diverges faster than the localization length. These results are in agreement
with those presented in (\cite{BCGD}); although the full correlations were
not computed exactly, it is shown in this paper that the two-point 
Green's function is indeed given by
\benn
\overline{\bra{x}\frac{1}{H-E}\ket{y}}\APPROX{E\to 0}\sum_n c_n 
\EXP{-\frac{\sigma\pi^2(2m+1)^2}{2\log^2E}|x-y|}
\eenn
It would be interesting to compare this method with Berezinski\u{i}'s
diagrammatic technique (\cite{berezinski2}) recently used in (\cite{steiner}).

\item The existence of a critical state at $E=0$ is best understood when
this model is reinterpreted in the context of classical diffusion in a
random medium. A diffusive behaviour at large time requires the existence of
an extended state at $E=0$ (\cite{tossatti}).

\item The thermodynamic properties of some one-dimensional spin systems with
random exchange couplings can be reinterpreted by using a mapping of the
spin system onto a model of free fermions. This approach, which can be traced
back to the pioneering work of Lieb, Schultz and Mattis (\cite{lieb}), was 
used  by Smith (\cite{smith})  in the context of the X-Y
model. Exact result for quantum phase transition in  random X-Y spin
chains with a comparison with the renormalization group approach
(\cite{fisher}) were obtained by McKenzie (\cite{mckenzie}). 
Quite recently, a similar approach with a
different type of disorder was developed by Fabrizio and M\'elin
(\cite{melin}). It is also
worth mentioning the nice paper of Steiner, Fabrizio and Gogolin
(\cite{steiner}) extending this analysis to the case of correlations and
boundary effects.

\end{enumerate}


\section{Correlation functions of the ground state wave function}
\label{correlation}

The localization properties of the wave function can be characterized by the
density-density correlation function. Various techniques have been
developed, mainly by the Russian school (\cite{lifpast}), to compute such
quantities in the weak disorder limit. In the supersymmetric model, one may
take advantage of the fact that the ground state wave function is known
exactly as a functional of the disorder. If the disordered potential
$\phi(x)$ is white noise,  this allows one to compute the
corresponding $n$-point function by using a mapping with Liouville quantum
mechanics. Such a calculation was recently carried out by Shelton and
Tsvelik (\cite{tsvelik}) for the case $n=2$ and $n=3$ in the context of spin
Peierls systems. An extension of this
result to arbitrary $n$ is given below. We first consider periodic boundary
conditions. For completness we also give the corresponding formula due to
Broderix and Kree in the case of free boundary conditions (\cite{broderix}). 

\subsection{Periodic boundary conditions}

We consider the supersymmetric Hamiltonian (\ref{susyham}) in which we
set $\frac{\beta U(x)}{2}\equiv\int_0^x \D x'\,\phi(x')$ for consistency with
the notation of previous work (\cite{fluxdist}, \cite{expfunc}). 
We are interested in 
the following two sections in the statistical properties of the zero mode 
wave function
\be\label{gs}
\psi_0(x)=
\frac{\EXP{\frac{\beta U(x)}{2}}}
     {\left[\int_0^L \D x'\,\EXP{\beta U(x')} \right]^{1/2}}
\ee
when the disordered potential $\phi(x)$ is white noise.
We consider a system of length $L$ and impose periodic boundary conditions.
This is achieved if the disordered potential is a Brownian bridge
($U(0)=U(L)$). The average over the disorder is performed through the Wiener
measure
\be
\mean{\cdots}={\cal N}\int_{U(0)=0}^{U(L)=0}{\cal D}U(x)\,\cdots
\EXP{-\frac{1}{2\sigma}\int_0^L \D x\,\left(\frac{\D U(x)}{\D x}\right)^2}
\ee
where ${\cal N}$ is a normalization to be determined.
Our aim is to compute the correlation functions of the square wave
function $\psi^2_0(x)$:
\be\label{Cdef}
C_n(x_1,\cdots,x_n)\equiv \mean{|\psi_0(x_1)|^2\cdots|\psi_0(x_n)|^2}
.\ee
In order to perform the average over $U(x)$ it is convenient to exponentiate 
the denominator coming from the normalization of the wave function. By
using the integral representation of the $\Gamma$ function, one gets
\beann
&&C_n(x_1,\cdots,x_n)\\
&&= \frac{\cal N}{\Gamma(n)}
\int_0^\infty \hspace{-0.3cm}\D p\,p^{n-1}
\int_{U(0)=0}^{U(L)=0}\hspace{-0.7cm}{\cal D}U(x)\,
\EXP{-\int_0^L \hspace{-0.2cm}\D x\,\left[
  \frac{1}{2\sigma}\left(\frac{\D U(x)}{\D x}\right)^2
  +p\EXP{\beta U(x)}\right]
+\beta U(x_1)+\cdots+\beta U(x_n)}
.\eeann
This expression establishes a link with one-dimensional Liouville quantum
mechanics. For a detailed discussion of this model we refer to
(\cite{fluxdist}) and (\cite{kolok2,kolok}).
Considerable simplification occurs if one makes use of the fact that, in
this theory, a change in the coupling constant can be interpreted as a
translation in $U$ space.
This suggests the change of variable $p=\alpha\EXP{\beta U}$, where
$\alpha\equiv\frac{\sigma\beta^2}{2}$, which leads to
\be
C_n(x_1,\cdots,x_n)=\frac{2\sqrt{\pi L}\,\alpha^{n+\frac{1}{2}}}{\Gamma(n)}
\int_{-\infty}^{+\infty}\D U\,
\int_{U(0)=U}^{U(L)=U}{\cal D}U(x)\,
\EXP{-S_{\rm L}}\EXP{\beta U(x_1)+\cdots+\beta U(x_n)}
\ee
where the action
\be
S_{\rm L}=\int_0^L \D x\,
\left[\frac{1}{2\sigma}\left(\frac{\D U(x)}{\D x}\right)^2
+\alpha\EXP{\beta U(x)}\right]
\ee
is associated with the Liouville Hamiltonian
\be
H_{\rm L}=-\frac{\sigma}{2}\frac{\D^2}{\D U^2} + \alpha \EXP{\beta U}
.\ee
The role of the normalization constant $\cal N$ is to insure the relation
\benn
\int\D x_1\cdots \D x_n\,C_n(x_1,\cdots,x_n)=1.
\eenn
Since $\cal N$ is independent
of $n$ this relation may be used in the case $n=1$ to find $\cal N$.

One may express the correlation function in terms of the Liouville propagator 
$G_x(U,U')\equiv\bra{U}\EXP{-xH_{\rm L}}\ket{U'}$. Choosing 
$L\geq x_1\geq x_2\geq \cdots \geq x_n\geq 0$ where
$x_{ij}\equiv x_i-x_j$ one gets
\beann
C_n(x_1,\cdots,x_n)&=&\frac{2\sqrt{\pi L}\,\alpha^{n+\frac{1}{2}}}{\Gamma(n)}
\int_{-\infty}^{+\infty}\D U_1 \D U_2\cdots \D U_n\,
G_{L-x_{1n}}(U_n,U_1)\EXP{\beta U_1} \\
& &\hspace{3cm} G_{x_{12}}(U_1,U_2)\EXP{\beta U_2}\cdots
G_{x_{n-1n}}(U_{n-1},U_n)\EXP{\beta U_n}\nonumber
.\eeann
Using the eigenstates $\ket{k}$ of the Liouville Hamiltonian\footnote{
The Liouville Hamiltonian has a continuous spectrum 
$H_{\rm L}\psi_k(U)=\frac{\alpha k^2}{4}\psi_k(U)$ where the wave function is
$\psi_k(U)=\frac{\sqrt{\beta k \sinh{\pi k}}}{\pi} 
K_{\I k}\left( 2\EXP{\frac{\beta U}{2}}\right)$.
} gives
\bea\label{corrme}
C_n(x_1,\cdots,x_n)&=&\frac{2\sqrt{\pi L}\,\alpha^{n+\frac{1}{2}}}{\Gamma(n)}
\int_0^{+\infty}\D k_1\cdots \D k_n\,
\bra{k_n}\EXP{\beta U}\ket{k_1}\cdots\bra{k_{n-1}}\EXP{\beta U}\ket{k_n}
\nonumber\\
& &\hspace{4.5cm}
\EXP{-\frac{\alpha(L-x_{1n})}{4}k_1^2-\cdots-\frac{\alpha x_{n-1n}}{4}k_n^2}
.\eea
Knowledge of the wave functions gives the matrix elements 
\be
\bra{k}\EXP{\beta U}\ket{k'}=
\frac{1}{8}\sqrt{kk'\sinh{\pi k}\sinh{\pi k'}}
\frac{k^2-k'^2}{\cosh{\pi k}-\cosh{\pi k'}}
.\ee
Expression (\ref{corrme}) allows one to get the long range behaviour (when all the
distances involved are large compared to $\frac{1}{\alpha}$). Using the Laplace
method one eventually finds
\be\label{res}
C_n(x_1,\cdots,x_n)\simeq\frac{1}{(4\pi\alpha)^\frac{n-1}{2}\Gamma(n)}\,
\frac{\sqrt{L}}{\left[(L-x_1+x_n)(x_1-x_2)\cdots (x_{n-1}-x_n)\right]^{3/2}}
.\ee
Despite the coefficients being different to those found by Shelton
and Tsvelik (\cite{tsvelik}) when $n=2$ and $n=3$, we get the same behaviour 
as a function of the distances.
This expression shows the existence of long range correlations.
A nice interpretation of the algebraic tail was recently given in
(\cite{ledoussal}). It is suggested that the exponent $3/2$ is associated with
configurations of the disorder where $U(x)$ returns to its starting point.
It is interesting to point out that the same exponent appears in conventional
localization theory. In this case  wave functions with the same
(\cite{berezinski2}, \cite{gogolin}) or nearly equal (\cite{gorkov}) energy are
weakly correlated at large distances.
The correlation function decays exponentially with the power law
preexponential factor $\left(\frac{l}{x}\right)^{3/2}$.

\subsection{Free boundary conditions}

Computation of correlation functions of the type in (\ref{Cdef}) was first
performed by Broderix and Kree (\cite{broderix}) in a different context:
$|\psi_0(x)|^2$ is interpreted as the equilibrium Gibbs measure for the
potential $U(x)$. Setting $U(0)=0$ and letting $U(L)$ be free they get

\be\label{resBK}
C^{\rm BK}_n(x_1,\cdots,x_n)\simeq
\frac{1}{\pi(4\pi\alpha)^\frac{n-1}{2} \Gamma(n)}
\frac{1}{\sqrt{x_n (L-x_1)}}\frac{1}{\left[
  (x_1-x_2)\cdots(x_{n-1}-x_n)
\right]^{3/2}}
\ee
which is very similar to (\ref{res}). The different boundary prescription
therefore induces a slight change in the $x$ dependance near the boundary.


\section{Multifractality}\label{moments}

In a recent series of publications, de Chamon et al. (\cite{chamon}) have 
considered a
two-dimensional localization  model which exhibits a localization transition
at zero energy. They consider a two-dimensional Dirac Hamiltonian in a 
random magnetic field $B(\vec r)=\Delta\phi(\vec r)$. 
In this case the zero energy solution
(\cite{A-C}) can be constructed explicitly for any realization of
$\phi(\vec r)$. They consider the particular ``ground state'' solution
$\psi(\vec r)\propto(\EXP{-\phi(\vec r)},0)$. For a Gaussian disorder
$P[\phi]=\exp{-\frac{1}{2g}\int\D \vec r\,(\vec\nabla\phi)^2}$
the successive moments of $\psi(\vec r)$ are encoded in the partition
function ${\cal Z}(q)\equiv\int\D \vec r\,\EXP{-2q\phi(\vec r)}$ since
$\int\D \vec r\,\psi^{2q}(\vec r)=\frac{{\cal Z}(q)}{{\cal Z}(1)^q}$. 
The multifractal exponents of the wave function can be obtained 
by using a formal equivalence with the problem of a directed polymer on the 
Cayley tree (\cite{derrida2}).

In the following we consider a one-dimensional version of this model with,
however, a different type of disorder and take into account the wave function
normalization as in (\cite{kogan}).
The starting point is the two-dimensional Euclidean Dirac operator
\benn
\I\Dslash=\I\sigma_1\left(\partial_y + \I A_y\right)
        + \I\sigma_2\left(\partial_x + \I A_x\right)
\eenn
$\sigma_i$ are the Pauli matrices and the gauge field is given by
\benn
\left\{\begin{array}{l}
A_y=f(x)\\
A_x=0.
\end{array}\right.
\eenn
We may take eigenstates of the form
$\psi(x,y)=\left( \chi(x)\EXP{i\omega y} ,0\right)$. The eigenvalue equation
$-\Dslash^2\chi=E^2\chi$ then becomes 
\benn
\left[-\partial_x^2 + (\omega+f(x))^2-\sigma_3
f'(x) \right]\chi(x)
=E^2\chi(x)
.\eenn
We are thus led to a one-dimensional Schr\"odinger equation with a
supersymmetric potential $V(x)=\varphi^2(x)-\sigma_3\varphi'(x)$ where 
$\varphi(x)=\omega+f(x)$.

In an earlier work by one of us (\cite{comtet0}) this approach was used to
study the density of states of the two-dimensional Dirac operator. If
$f(x)$ is white noise we may use the density of states of the
one-dimensional problem and integrate over the free motion on the $y$
axis. The resulting expression displays an enhancement at low energy as
compared to the free case.

Here our purpose is to characterize the fluctuation properties of the ground
state wave function  $\chi(x)= \EXP{-\int\D y\varphi(y)}$. 
By using two different approaches for two boundary prescription we compute 
exactly the successive moments of the normalized ground state
\benn
\psi_0(x)=\frac{\chi(x)}{\left[\int_0^L \D x'\,\chi^2(x')\right]^{1/2}}
\eenn
In order to keep unified conventions we will parametrize $\psi_0(x)$ as in
equation (\ref{gs}).

\subsection{Moments of $\psi_0(x)$ for periodic boundary conditions}

If the random potential obeys periodic boundary conditions, the situation is
the one described in section \ref{correlation}. The moment of order $2n$ is
then related to the $n$-point correlation function at coinciding points
\benn
\mean{|\psi_0(x)|^{2n}}=C_n(x,\cdots,x)
.\eenn
In terms of the Liouville propagator this is expressed as
\beann
\mean{|\psi_0(x)|^{2n}}&=&
\frac{2\sqrt{\pi L}\,\alpha^{n+\frac{1}{2}}}{\Gamma(n)}
\int_{-\infty}^{+\infty}\D U_1\, G_L(U_1,U_1)\EXP{n\beta U_1}\\
&=& \frac{2\sqrt{\pi L}\,\alpha^{n+\frac{1}{2}}}{\Gamma(n)}
\int_0^\infty \D k\,\bra{k}\EXP{n\beta U}\ket{k}\EXP{-\frac{\alpha L}{4}k^2}
\eeann
where the matrix element is
\benn
\bra{k}\EXP{n\beta U}\ket{k}=\frac{\Gamma(n)^2}{4\pi\Gamma(2n)}
\prod_{m=0}^{n-1} \left(m^2+k^2\right)
.\eenn
The moments are eventually given by
\be
\mean{|\psi_0(x)|^{2n}}=\frac{\alpha^n}{2^n(2n-1)!!}
\sum_{m=1}^n a_m^n 2^m(2m-1)!! \frac{1}{(\alpha L)^m}
\ee
where the coefficients $a_m^n$ are defined by the equation
\benn
\prod_{m=0}^{n-1} \left(m^2+X\right)=\sum_{m=1}^n a_m^n X^m.
\eenn
For example $a_n^n=1$ and $a_1^n=\Gamma(n)^2$.

One may extract from this expression the asymptotic dependance of the moments
(when $L$ is large compared to $\frac{1}{\alpha}$)
\be\label{mompb}
\mean{|\psi_0(x)|^{2n}}\simeq
\left(\frac{\alpha}{2}\right)^{n-1}
\frac{\Gamma(n)^2}{(2n-1)!!}\,\frac{1}{L}
.\ee

\subsection{Moments of $\psi_0(x)$ for free boundary conditions}

If one leaves $U(L)$ free instead of imposing periodic boundary, then the 
Brownian that enters in the wave function (\ref{gs}) is no longer a Brownian 
bridge but a free Brownian motion starting from $U(0)=0$.

In the average over the disordered potential one must now take into account all 
the Brownian paths starting from $0$ without any restriction on the final 
point. Except for this slight modification the formalism is the same as in
previous sections.
One is led to
\beann
\mean{\psi_0^{2n}(x_1)}&=&
\frac{\beta\alpha^n}{\Gamma(n)}\int_{-\infty}^{+\infty} \D U
\int_{U(0)=U} {\cal D}U(x)\,\EXP{-S_{\rm L}}\EXP{n\beta U(x_1)}\\
&=&\frac{\beta\alpha^n}{\Gamma(n)}\int_{-\infty}^{+\infty}\D U \D U_1 \D U_2\,
G_{L-x}(U_2,U_1)\EXP{n\beta U_1}G_x(U_1,U)
.\eeann

It is interesting to present another derivation of the moments using 
the language of
previous works (\cite{oshanin}, \cite{fluxdist}, \cite{expfunc}) devoted 
to the study of the statistical properties of the exponential functional 
$Z_L^{(\mu)}\equiv\int_0^L \D x\,\EXP{-(\mu\alpha x+\sqrt{2\alpha}W(x))}$.
Here $\mu$ is the drift for the Brownian motion
$\beta U(x)=-(\alpha\mu x+\sqrt{2\alpha} W(x))$ and $W(x)$ is a Brownian
motion of average $0$ and variance $1$ defined on $[0,L]$.

We will prove that the moments of $\psi_0(x)$ can be expressed in terms of
the characteristic function
$\phi^{(\mu)}(p,L)\equiv\langle\EXP{-pZ_L^{(\mu)}}\rangle$ (Laplace
transform of the distribution law of $Z_L^{(\mu)}$).
For this purpose we may note that $\psi_0(x)$ involves two
exponential functionals. We may write
\benn
\psi_0^2(x)=\frac{1}{\int_0^L \D y\, \EXP{\beta (U(y)-U(x))}}
\eenn
and separate the denominator into two parts
\benn
\int_0^L \D y\, \EXP{\beta (U(y)-U(x))}=
\int_0^x \D y\, 
   \EXP{-[-\mu\alpha y+\sqrt{2\alpha}B(y)]}
+\int_0^{L-x} \D y\, 
   \EXP{-[\mu\alpha y+\sqrt{2\alpha}\tilde B(y)]}
\eenn
where $B(y)\equiv W(x-y)-W(x)$ and $\tilde B(y)\equiv W(x+y)-W(x)$,
respectively defined on $y\in[0,x]$ and $y\in[0,L-x]$, are two independent
Brownian motions starting from zero $B(0)=\tilde B(0)=0$. The denominator is
then a sum of two statistically independent exponential functionals 
$Z_x^{(-\mu)}$ and $\tilde Z_{L-x}^{(\mu)}$.

The moments may then be rewritten as
\beann
\mean{\psi_0^{2n}(x)}&=&
\mean{\frac{1}{\left( Z_x^{(-\mu)} + \tilde Z_{L-x}^{(\mu)} \right)^n}}\\
&=&\frac{1}{\Gamma(n)}\int_0^\infty \D p\,p^{n-1} 
\mean{\EXP{-p Z_x^{(-\mu)}}}
\mean{\EXP{-p\tilde Z_{L-x}^{(\mu)}}}\\
&=&\frac{1}{\Gamma(n)}\int_0^\infty \D p\,p^{n-1} \phi^{(-\mu)}(p,x)
\:\phi^{(\mu)}(p,L-x)
.\eeann
Where the characteristic functions are given in
(\cite{fluxdist}, \cite{expfunc}).
In the case $\mu=0$ which is of interest for us 
\be\label{phip}
\phi^{(0)}(p,L)=\frac{2}{\pi}\int_0^\infty \D s\,
\cosh{\frac{\pi s}{2}}\:K_{\I s}\left( 2\sqrt{\frac{p}{\alpha}}\right) 
\EXP{-\frac{\alpha L}{4} s^2}
.\ee
One may extract from this expression the dominant behaviour when 
$\alpha x\gg 1$ and  $\alpha(L-x)\gg 1$
\be
\mean{\psi_0^{2n}(x)}\simeq\left(\frac{\alpha}{2}\right)^{n-1}
\frac{\Gamma(n)^2}{\pi(2n-1)!!}\frac{1}{\sqrt{x(L-x)}}
.\ee
One may check that this result agrees with the one of Broderix and Kree 
(\ref{resBK}) in the case $n=1$.
Whenever $x$ belongs to the interval $[0,L]$, but is not on the edges,
the moments still behave as $\mean{\psi_0^{2n}(x)}\sim\frac{1}{L}$. 

We may also explore the behaviour of the moments of the wave function on the
edges $\mean{\psi_0^{2n}(0)}=\mean{\psi_0^{2n}(L)}$. This quantity is equal
to the moments of the partition function $Z_L^{(0)}$ for negative orders
\benn
\mean{\psi_0^{2n}(0)}=\mean{\frac{1}{(Z_L^{(0)})^n}}
=\frac{1}{\Gamma(n)}\int_0^\infty \D p\, p^{n-1} \phi^{(0)}(p,L)
.\eenn
We may easily extract the dominant behaviour of this quantity with the
help of (\ref{phip}) and eventually find 
\be\label{mom}
\mean{\psi_0^{2n}(0)}\simeq
\frac{\alpha^n\Gamma(n)}{\sqrt{\pi\alpha L}}
.\ee
This shows that the wave function fluctuates more on the edges of the
interval (when $x\ll\frac{1}{\alpha}$ or $L-x\ll\frac{1}{\alpha}$) than in
the bulk (when $x\gg\frac{1}{\alpha}$ and $L-x\gg\frac{1}{\alpha}$).

The square of the wave function on the edge is interpreted in the problem of
classical diffusion in the random potential $U(x)$, as the steady current
density when a constant density of particle is imposed at $x=0$ in the
presence of a trap at $x=L$.
The distribution of the steady current was found in (\cite{oshanin2}, 
\cite{fluxdist}). 
Setting $J\equiv\frac{1}{\alpha}\psi_0^2(0)$ the distribution of $J$ is a 
log-normal law for small values of $J$
\benn
P(J)\APPROX{J\to 0} 
\frac{1}{2\sqrt{\pi\alpha L}}\EXP{-\frac{1}{4\alpha L}\log^2 J}
\eenn
and possesses an exponential tail for large values of $J$
\benn
P(J)\APPROX{J\to \infty} 
\frac{1}{\sqrt{\pi\alpha L}}\:\frac{1}{J}
\EXP{\frac{\pi^2}{4\alpha L}-J}
\eenn
which is responsible of the behaviour of the moments given in (\ref{mom}).

\subsection{Discussion of the results}

We have just seen that the one-dimensional model is closely related to
the model studied in (\cite{chamon}). Since the one-dimensional model is
exactly solvable, it is interesting to study its multifractal properties,
although one cannot expect it to give the same behaviour. In fact, in both 
models the potential is long-range correlated; though in $d=2$ the 
correlations are logarithmic whereas in $d=1$ they are linear.

We define the scaling exponent $\tilde\tau(q)$ which characterizes the 
behaviour of the critical wave function at $E=0$ as
\benn
\mean{|\psi(\vec r)|^{2q}}\APPROXs{L\to\infty}L^{-d-\tilde\tau(q)}
.\eenn
Using equation (\ref{mompb}) we get $\tilde\tau(q)=0$. This exponent is
the scaling exponent of the average moments of the wave function.
We may introduce a slightly different scaling exponent
\benn
\tau(q)=\lim_{L\to\infty}
\frac{\mean{\log \int\D \vec r\,|\psi_0(\vec r)|^{2q}}}{\log (1/L)}
,\eenn
which is the mean exponent of the critical wave function.
In terms of the partition function 
$Z^{(0)}_L(\beta)=\int_0^L \D x\,\EXP{\beta U(x)}$ 
introduced before, it may be rewritten
\benn
\tau(q)=\lim_{L\to\infty}
\frac{\mean{\log Z^{(0)}_L(q\beta)}-q\mean{\log Z^{(0)}_L(\beta)}}{\log (1/L)}
.\eenn
The mean free energy that appears has been calculated in (\cite{expfunc}) to
be
\benn
\mean{\log Z^{(0)}_L(\beta)}
=2\sqrt{\frac{\alpha L}{\pi}}+C-\log\alpha -\frac{\pi}{3\sqrt{\alpha L}}
+{\rm O}\left(\frac{1}{(\alpha L)^{3/2}}\right)
,\eenn
where $C=-\Gamma'(1)$ is the Euler-Mascheroni constant.
Again, one eventually gets a scaling exponent $\tau(q)=0$
which agrees with the
behaviour found in (\cite{chamon}) in the strong disorder regime. By using the
mapping with the random directed polymer model (\cite{derrida2}), de Chamon
et al. have shown that this regime corresponds in fact to the low 
temperature phase of this model.
The existence of such a link also appears in the one-dimensional case. A
comparison of the moments of $Z^{(0)}_L(\beta)$ with those of the random
energy model (REM) (\cite{derrida}) reveals some striking similarities.
For the one-dimensional case the expression of the moments given
in (\cite{fluxdist}) read
\benn
\mean{(Z_L^{0})^n}=\frac{1}{\alpha^n}
\left( \frac{\Gamma(n)}{\Gamma(2n)} 
     \sum_{k=1}^n (-1)^{n-k}\EXP{\alpha L k^2} C_{2n}^{k+n}   
   + \frac{(-1)^n}{n!}\right)
.\eenn
In particular, for large $L$ one obtains
\benn
\mean{(Z_L^{0})^n}\APPROX{\alpha L\gg 1}\frac{1}{\alpha^n}
\frac{\Gamma(n)}{\Gamma(2n)} \EXP{\alpha L n^2}
.\eenn
These moments grow in the same manner as in the REM. The main difference is
that, unlike in the REM, there is here no transition above which the
behaviour would change from $\EXP{\alpha L n^2}$ to $\EXP{\alpha L n}$. In
some sense one can consider that the transition temperature is sent to 
infinity. This explains why, in this one-dimensional case, one only probes
the low temperature phase. It would be extremely interesting to explore
intermediate cases with weaker correlations, as recently suggested in
(\cite{mezard}).


\section*{Acknowledgments}

Many of these results are based on earlier works done in collaboration with
Jean-Philippe Bouchaud, Jean Desbois, Pierre Le~Doussal, Antoine Georges,
Gleb Oshanin, C\'ecile Monthus and Marc Yor. A great thanks to all of them.
We especially thank C\'ecile Monthus for drawing our attention to the recent
literature and for interesting remarks. We thank Gleb Oshanin and Serguei
Nechaev for interesting discussion.



\begin{thebibliography}

\bibitem{}{A-C}{Aharonov and Casher 1979}
Aharonov, Y., Casher, A. (1979):
Ground state of a spin-1/2 charged particle in a two-dimensional magnetic
field,
Phys. Rev. A {\bf 19}(6), 2461--2462.

\bibitem{}{balents}{Balents and Fisher 1997}
Balents, L., Fisher, M. P. (1997):
Delocalization transition via supersymmetry in one dimension,
preprint , cond-mat/9706069.

\bibitem{}{berezinski2}{Berezinski\u{i} 1974}
Berezinski\u{i}, V. (1974):
Kinetics of a quantum particle in a one-dimensional random potential,
Sov. Phys. JETP {\bf 38}(3), 620.

\bibitem{}{pastur}{Benderski\u{i} and Pastur 1974}
Benderski\u{i}, M., Pastur, L. (1974):
Sov. Phys. JETP {\bf 40}, 241.

\bibitem{}{BCGD2}{Bouchaud et al. 1987}
Bouchaud, J.-P., Comtet, A., Georges, A., Le~Doussal, P. (1987):
Europhys. Lett. {\bf 3}, 653.

\bibitem{}{BCGD}{Bouchaud  et al. 1990}
Bouchaud, J.-P., Comtet, A., Georges, A., Le~Doussal, P. (1990):
Classical diffusion of a particle in a one-dimensional random force field,
Ann. Phys. (N.Y.) {\bf 201}, 285--341.

\bibitem{}{classdiff}{Bouchaud and Georges 1990}
Bouchaud, J.-P., Georges, A. (1990):
Anomalous diffusion in disordered media: Statistical mechanisms, models and
physical application,
Phys. Rep. {\bf 195}, 267.

\bibitem{}{mezard}{Bouchaud and M\'ezard 1997}
Bouchaud, J.-P., M\'ezard, M. (1997):
Universality classes for extreme value statistic,
preprint , cond-mat/9707047.

\bibitem{}{bovier}{Bovier 1989}
Bovier, A. (1989):
J. Stat. Phys. {\bf 56}, 645.

\bibitem{}{broderix}{Broderix and Kree 1995}
Broderix, K., Kree, R. (1995):
Thermal equilibrium with the Wiener potential: testing the replica 
variational approximation,
Europhys. Lett. {\bf 32}, 343--348.

\bibitem{}{casherneu}{Casher et al. 1984}
Casher, A., Neuberger, H. (1984):
Phys. Rev. B {\bf 139}, 67.

\bibitem{}{chamon}{de Chamon et al. 1996}
de~C.~Chamon, C., Mudry, C., Wen, X.-G. (1996):
Localization in two dimensions, Gaussian field theory and Multifractality,
Phys. Rev. Lett. {\bf 77}, 4194--4197.

\bibitem{}{comtet0}{Comtet et al. 1988}
Comtet, A., Georges, A., Le~Doussal, P. (1988):
Exact density of states of a two-dimensional Dirac operator in a random 
magnetic field,
Phys. Lett. B {\bf 208}(3/4), 487.

\bibitem{}{locprop}{Comtet et al. 1995}
Comtet, A., Desbois, J., Monthus, C. (1995):
Localization properties in one-dimensional disordered supersymmetric quantum 
mechanics,
Ann. Phys. (N.Y.) {\bf 239}, 312--350.

\bibitem{}{expfunc}{Comtet et al. 1996}
Comtet, A., Monthus, C., Yor, M. (1996):
Exponential functionals of Brownian motion and disordered systems,
to appear in J. of Appl. Prob. , preprint cond-mat/9601014.

\bibitem{}{cooper}{Cooper et al. 1995}
Cooper, F., Khare, A., Sukhatme, U. (1995):
Supersymmetry and quantum mechanics,
Phys. Rep. {\bf 251}, 267--385.

\bibitem{}{derrida}{Derrida 1981}
Derrida, B. (1981):
Random-energy model: an exactly solvable model of disordered systems,
Phys. Rev. B {\bf 24}(5), 2613--2626.

\bibitem{}{derrida2}{Derrida and Spohn 1988}
Derrida, B., Spohn, H. (1988):
Polymers on disordered trees, spin glasses and traveling waves,
J. Stat. Phys. {\bf 51}, 817.

\bibitem{}{dyson}{Dyson 1953}
Dyson, F. J. (1953):
Phys. Rev. {\bf 92}, 1331.

\bibitem{}{eggarter}{Eggarter and Riedinger 1978}
Eggarter, T., Riedinger, R. (1978):
Phys. Rev. {\bf 18}, 569.

\bibitem{}{melin}{Fabrizio and M\'elin 1997}
Fabrizio, M., M\'elin, R. (1997):
Coexistence of antiferromagnetism and dimerization in a disordered 
spin-Peierls model: exact results,
preprint , cond-mat/9701149.

\bibitem{}{fisher}{Fisher 1994}
Fisher, D. (1994):
Phys. Rev. B {\bf 50}(6), 3799.

\bibitem{}{stern}{Floratos et al. 1980}
Floratos, E., Stern, J. (1980):
Phys. Lett. B {\bf 119}, 419.

\bibitem{}{gogolin}{Gogolin 1975}
Gogolin, A., Melnikov, V., Rashba, E. (1975):
Zh. Eksp. Teor. Fiz. {\bf 69}, 328.

\bibitem{}{gorkov}{Gor'kov et al. 1983}
Gor'kov, L.,  Dorokhov, O., Prigara, F. (1983):
Sov. Phys. JETP {\bf 58}(4), 852.

\bibitem{}{kogan}{Kogan et al. 1996}
Kogan, I. I., Mudry, C., Tsvelik, A. M. (1996):
Liouville Theory as a model for prelocalized states in disordered conductors,
Phys. Rev. Lett. {\bf 77}(4), 707.

\bibitem{}{kolok2}{Kolokolov 1993}
Kolokolov, I. (1993):
The method of functional integration for a one-dimensional
localization, higher correlators and the average current flowing in a
mesoscopic ring in an arbitrary magnetic field,
Sov. Phys. JETP {\bf 76}, 1099.

\bibitem{}{kolok}{Kolokolov 1994}
Kolokolov, I. (1994):
Europhys. Lett. {\bf 28}, 193.

\bibitem{}{ledoussal}{Laloux and Le~Doussal 1997}
Laloux, L., Le~Doussal, P. (1997):
Aging and diffusion in low dimensional environments,
preprint , cond-mat/9705249.

\bibitem{}{lieb}{Lieb et al. 1961}
Lieb, E., Schultz, T., Mattis, D. (1961):
Ann. Phys. (N.Y.) {\bf 16}, 407.

\bibitem{}{lifshits}{Lifshits 1965}
Lifshits, I. M. (1965):
Energy spectrum structure and quantum states of disordered condensed systems,
Sov. Phys. Usp. {\bf 18}(4), 549.

\bibitem{}{lifpast}{Lifshits et al. 1988}
Lifshits, I. M., Gredeskul, S. A., Pastur, L. A. (1988):
{\em Introduction to the theory of disordered systems},
John Wiley \& Sons.

\bibitem{}{luck}{Luck 1992}
Luck, J.-M. (1992):
{\em Syst\`emes d\'esordonn\'es unidimensionnels},
Al\'ea Saclay.

\bibitem{}{luttinger}{Luttinger and Waxler 1987}
Luttinger, J., Waxler, R. (1987):
Ann. Phys. (N.Y.) {\bf 175}, 319.

\bibitem{}{mckenzie}{McKenzie 1996}
McKenzie, R. H. (1996):
Phys. Rev. Lett. {\bf 77}(23), 4804.

\bibitem{}{markos}{Markos 1988}
Markos, P. (1988):
Z. Phys. B - Cond. Matt. {\bf 73}, 17.

\bibitem{}{fluxdist}{Monthus and Comtet 1994}
Monthus, C., Comtet, A. (1994):
On the flux distribution in a one-dimensional disordered system,
J. Phys. I (France) {\bf 4}, 635--653.

\bibitem{}{gsenergy}{Monthus et al. 1996}
Monthus, C., Oshanin, G., Comtet, A., Burlatsky, S. (1996):
Sample-size dependence of the ground-state energy in a
one-dimensional localization problem,
Phys. Rev. E {\bf 54}, 231--242.

\bibitem{}{oshanin}{Oshanin et al. 1993a}
Oshanin, G., Mogutov, A., Moreau, M. (1993a):
Steady flux in a continuous-space Sinai chain,
J. Stat. Phys. {\bf 73}, 379.

\bibitem{}{oshanin2}{Oshanin et al. 1993b}
Oshanin, G., Burlatsky, S., Moreau, M., Gaveau, B. (1993b):
Behaviour of transport characteristics in several one-dimensional disordered
systems,
Chem. Phys. {\bf 177}, 803.

\bibitem{}{ovchinni}{Ovchinnikov and Erikmann 1977}
Ovchinnikov, A., Erikmann, N.  (1977):
Zh. Eksp. Teor. Fiz. {\bf 73}, 650.

\bibitem{}{neuberger}{Neuberger 1982}
Neuberger, H. (1982):
Phys. Lett. B {\bf 112}(4/5), 341.

\bibitem{}{tsvelik}{Shelton and Tsvelik 1997}
Shelton, D. G., Tsvelik, A. M. (1997):
An effective theory for midgap states in doped spin ladder and spin-Peierls 
systems: Liouville quantum mechanics,
preprint , cond-mat/9704115.

\bibitem{}{smith}{Smith 1970}
Smith, E. (1970):
J. Phys. C {\bf 3}, 1419.

\bibitem{}{steiner}{Steiner et al. 1997}
Steiner, M.,  Fabrizio, M., Gogolin, A. O. (1997):
Random mass Dirac fermions in doped spin-Peierls and spin-ladder
systems: one-particle properties and boundary effects,
preprint , cond-mat/9706096.

\bibitem{}{theodorou}{Theodorou and Cohen 1976}
Theodorou, G., Cohen, M. (1976):
Phys. Rev. B {\bf 13}, 4597.

\bibitem{}{tossatti}{Tossatti 1990}
Tossatti, E., Vulpiani, A., Zannetti, M. (1990):
Physica A {\bf 164}, 705.

\bibitem{}{yor3}{Yor 1992}
Yor, M. (1992):
Adv. Appl. Prob. {\bf 24}, 509.

\bibitem{}{witten}{Witten 1981}
Witten, E. (1981):
Nucl. Phys. B [FS] {\bf 188}, 513.

\end{thebibliography}
\end{document}